\documentclass[%
 reprint,
superscriptaddress,
showpacs,
 amsmath,amssymb,
 aps, longbibliography
]{revtex4-1}

\usepackage{rotating}
\usepackage{graphicx}
\usepackage{color}
\usepackage{verbatim}
\usepackage{subfigure}
\usepackage{tkz-graph}

\DeclareMathOperator{\diag}{diag}
\definecolor{BlueB}{rgb}{0,0,1}
\definecolor{Red}{rgb}{1,0,0}
\definecolor{Green}{rgb}{0,0.5,0}
\definecolor{Black}{rgb}{0,0,0}
\newcommand{\comments}[1]{{\color{Black} #1}}
\newcommand{\NewText}[1]{{\color{Black} #1}}
\newcommand{\commentsB}[1]{{\color{Black} #1}}

\newcommand{\David}[1]{{\color{Black} #1}}
\newcommand{\Da}[1]{{\color{Black} #1}}
\newcommand{\MOD}[1]{{\color{Black} #1}}
\newcommand{\Final}[1]{{\color{Black} #1}}

\begin{document}

\preprint{APS/123-QED}

\title{Versatile photonic entanglement synthesizer in the spatial domain} 

\author{David Barral} 
\email{david.barral@universite-paris-saclay.fr}
\affiliation{Centre de Nanosciences et de Nanotechnologies C2N, CNRS, Universit\'e Paris-Saclay, 10 boulevard Thomas Gobert, 91120 Palaiseau, France}
\author{Mattia Walschaers}
\affiliation{Laboratoire Kastler Brossel, Sorbonne Universit\'e, CNRS, ENS-PSL Research University, Coll\`ege de France, 4 place Jussieu, F-75252 Paris, France}
\author{Kamel Bencheikh}
\affiliation{Centre de Nanosciences et de Nanotechnologies C2N, CNRS, Universit\'e Paris-Saclay, 10 boulevard Thomas Gobert, 91120 Palaiseau, France}
\author{Valentina Parigi}
\affiliation{Laboratoire Kastler Brossel, Sorbonne Universit\'e, CNRS, ENS-PSL Research University, Coll\`ege de France, 4 place Jussieu, F-75252 Paris, France}
\author{Juan Ariel Levenson}
\affiliation{Centre de Nanosciences et de Nanotechnologies C2N, CNRS, Universit\'e Paris-Saclay, 10 boulevard Thomas Gobert, 91120 Palaiseau, France}
\author{Nicolas Treps}
\affiliation{Laboratoire Kastler Brossel, Sorbonne Universit\'e, CNRS, ENS-PSL Research University, Coll\`ege de France, 4 place Jussieu, F-75252 Paris, France}
\author{Nadia Belabas}
\email{nadia.belabas@universite-paris-saclay.fr}
\affiliation{Centre de Nanosciences et de Nanotechnologies C2N, CNRS, Universit\'e Paris-Saclay, 10 boulevard Thomas Gobert, 91120 Palaiseau, France}

\begin{abstract}
Multimode entanglement is an essential resource for quantum information in continuous-variable systems. Light-based quantum technologies will arguably not be built upon table-top bulk setups, but will presumably rather resort to integrated optics. Sequential bulk optics-like proposals based on cascaded integrated interferometers are not scalable with the current state-of-the-art low-loss materials used for continuous variables. We analyze the multimode continuous-variable entanglement capabilities of a compact currently-available integrated device without bulk-optics analog: the array of nonlinear waveguides. We \Da{theoretically} demonstrate that this simple and compact structure, together with a reconfigurable input pump distribution and multimode coherent detection of the output modes, is a versatile entanglement synthesizer in the spatial domain. We exhibit this versatility through analytical and numerically optimized multimode squeezing, entanglement, and cluster state generation in different encodings. Our results re-establish spatial encoding as a contender in the game of continuous-variable quantum information processing.

\end{abstract}

\date{October 12, 2020}
\maketitle 

\section{Introduction}\label{I}

Two key phenomena underpin current quantum technologies: quantum superposition and quantum correlations --entanglement-- \cite{Acin2018}. The paradigmatic example of entanglement is the case of two spatially separated quantum particles that have both maximally correlated momenta and maximally anticorrelated positions \cite{Reid2009}. Position and momentum are continuous variables (CV), i.e. variables that take a continuous spectrum of eigenvalues. In the optical domain CV-based quantum information can be encoded in the fluctuations of the electromagnetic field quadratures \cite{Furusawa2015}. Features like deterministic resources, unconditional operations and near-unity efficiency homodyne detectors make CV a powerful framework for the development of quantum technologies \cite{Braunstein2005}. Remarkably, entanglement between more than two parties is also useful. Large-scale CV entangled states are the resources of a promising class of quantum computing: measurement-based quantum computing (MBQC) \cite{Raussendorf2001, Menicucci2006}. A recent breakthrough generating large two-dimensional cylindrical-array clusters positions CV as a frontrunner in the race after a photonic quantum computer \cite{Larsen2019, Asavanant2019}. Multipartite entangled states are up to now produced in table-top experiments with specific designs generating only specific entanglement geometries or quantum networks: recent demonstration of large-scale entanglement include frequency \cite{Chen2014}, temporal \cite{Yoshikawa2016} and spatial \cite{Yukawa2008, Su2013, Armstrong2015} encoding of squeezed light.

The on-demand generation of different multimode entangled states with the same optical setup --an entanglement synthesizer-- is a challenging task. Entanglement synthesizers \NewText{are key for applications} in quantum computing and quantum simulation \cite{Nokkala2018, Killoran2019}. Transverse laser modes and frequency modes entanglement synthesizers based respectively on postprocessing measurement results and on measurement basis shaping have been introduced in \cite{Armstrong2012, Roslund2013, Cai2017}. \Final{These supermode-based approaches are interesting for engineering quantum states and simulating complex quantum systems}. They are however not fully equivalent to a quantum network, since the quantum information can only be processed locally, but not in a distributed way. Recently, an entanglement synthesizer with the possibility to distribute the nodes has been introduced in the time domain \cite{Takeda2019}. A versatile entanglement synthesizer in the spatial domain is nevertheless still missing. Specifically, \Final{spatial} distribution of signal in networks is naturally accomplished in this domain \Da{and this is highly desirable for applications} in secure communications via quantum secret sharing \cite{Hillery1999} or in distributed quantum sensing \cite{Guo2019}. In this work we introduce the monolithic array of waveguides with built-in nonlinearity as a versatile spatial-mode squeezing and entanglement synthesizer.

 \begin{figure*}[t]
  \centering
    \includegraphics[width=0.98\textwidth]{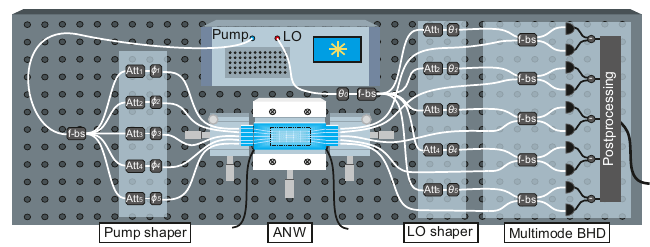}
\vspace {0cm}\, 
\hspace{0cm}\caption{\label{F1}\small{\Da{Sketch of our versatile entanglement synthesizer based on an array of nonlinear waveguides (ANW) made up of five PPLN waveguides working in a SPDC configuration (not at scale). A single laser with a second harmonic generation stage outputs coherent pump and local oscillator (LO) beams. A reconfigurable multimode shaper composed by a $1\times N$ fiber beam splitter (f-bs), attenuators (Att$_{j}$) and phase shifters ($\phi_{j}$) at pump frequency inputs the desired profile ($\vec{\eta}, \vec{\phi}$) in the array through a V-groove array. Bent waveguides conduct the pump modes to the periodically poled ANW (dashed box) where signal modes are generated and evanescently coupled. The coupling profile $\vec{f}$, wavevector phasematching and coupling phasematching can be suitably engineered for a specific operation mode. The output light is collected by V-groove arrays and directed to a multimode balanced homodyne detector (BHD). The multimode LO shaper is composed by attenuators (Att$_{j}$) and phase shifters ($\theta_{j}$). Each independent BHD mixes LO and SPDC light in a balanced f-bs and the result is measured by a pair of photodiodes. The currents yielded by photodiodes are subtracted two by two and the generated electric signals are then electronically mixed (postprocessed) if necessary. Access to the individual mode basis is achieved simply using independent LOs. Access to the supermode bases involves shaping the LO in every single supermode through LO phase ($\vec{\theta}$) and amplitude profiles (see sections \ref{IIIb} and \ref{Vb}).
}}}
\end{figure*}

We propose to take advantage of integrated optics, a solution for the blooming of market-scalable quantum-optics technologies \cite{Wang2019}. Spatial encoding is a natural framework in integrated optics where the information is encoded in the propagating waveguided modes and transmitted by optical fibers. A source of CV entangled states fully on-chip has been introduced recently \cite{Lenzini2018}. However, the extension of that scheme of sequential squeezing and entanglement --via nonlinear waveguides and directional couplers, respectively-- to \Da{a number of modes $N>2$} is very demanding with current technology. We establish here that the array of nonlinear waveguides (ANW) \Da{ --that interlaces nonlinearity and \Final{evanescent} coupling, i.e. interlaces squeezing and entanglement, in a monolithic way (Figure \ref{F1})--} is a good contender for versatile and scalable generation of entanglement. CV entanglement generation in monolithic ANW is \Da{thus} paradigmatically different from the current sequential bulk-optics-inspired schemes \cite{Yukawa2008, Su2013, Armstrong2015}. \Da{Nonclassical biphoton states for applications in the discrete-variable (DV) regime have been demonstrated in such waveguide arrays \cite{Kruse2013, Solntsev2014}. On-chip DV cluster states have been recently demonstrated and proposed in a bulk-optics-inspired scheme and the ANW, respectively \cite{Adcock2019, Titchener2020}.} Bipartite and tripartite CV entanglement have been predicted in arrays of nonlinear waveguides in the spontaneous (SPDC) and stimulated parametric downconversion regime \cite{Rai2012}. Nevertheless, tripartite entanglement was only predicted there for critical values of the involved parameters. We have proposed recently a scalable approach to generate multipartite entanglement in arrays of nonlinear waveguides in the second harmonic generation regime \cite{Barral2019b}. We show here that, in the experimentally relevant regime of SPDC, reconfigurable multimode pump and measurement, and freedom of choice of basis encoding enable on-demand programmability of entanglement. We \Da{theoretically} demonstrate the versatility of our approach in the production of multimode squeezing, multipartite entanglement and cluster states, through analytical and numerical solutions, applying optimization strategies. Remarkably, since the encoding of quantum information is mode basis-dependent, we present two cluster-state-generation procedures: i) entanglement generation among individual modes of the waveguides, and ii) \Final{entanglement generation among the elements of optimized nonlocal bases of the array}. \Da{We thus demonstrate that our photonic entanglement synthesizer can operate in the two regimes pioneered in free space CV respectively in i) \cite{Yukawa2008, Su2013, Armstrong2015} and ii) \cite{Armstrong2012, Roslund2013, Cai2017}. We review the possible applications in each case.} 

\Da{The article is organized as follows: in section \ref{II} we introduce the \Final{theoretical model for the} ANW. In section \ref{III} we develop and solve the propagation equations in three relevant encoding bases: individual modes, linear and nonlinear supermodes. We propose methods of detection in each basis. We then exhibit in all bases unoptimized generation of multimode squeezing using one analytical solution related to a specific pumping profile. We conclude summarizing the main features of each encoding basis. In section \ref{IV} we analyze the optimized generation of multipartite entanglement in the individual mode basis. In section \ref{V} we focus on the versatile generation of cluster states both in terms of their geometry and encoding. Finally, in section \ref{VI} we discuss on the different mode-basis encoding and point at applications, we present future research directions for our synthesizer, and we analyze the feasibility of our approach.}

\section{The array of nonlinear waveguides}\label{II}

The array of nonlinear waveguides consists of $N$ identical \Da{single-mode} $\chi^{(2)}$ waveguides in which degenerate SPDC and \Da{nearest-neighbor} evanescent coupling between the generated fields take place. The array can be made up of, for instance, periodically poled lithium niobate (PPLN) waveguides as sketched in Figure \ref{F1} \MOD{(dashed box)}. In each waveguide, an input harmonic field at frequency $\omega_{h}$ is type-0 downconverted (same polarization modes) into a signal field at frequency $\omega_{s}$  \cite{Alibart2016}. We consider pump undepletion with $\alpha_{h,j}$ a strong coherent pump field propagating in the $j$th waveguide. We consider that the phase matching condition  $\Delta\beta\equiv\beta(\omega_{h})-2\beta(\omega_{s})=0$, with $\beta(\omega_{h,s})$ the propagation constant at frequency $\omega_{h,s}$, is fulfilled only in the coupling zone. The energy of the signal modes propagating in each waveguide is exchanged between the coupled waveguides through evanescent waves, whereas the interplay of the second harmonic waves is negligible for the considered propagation lengths due to their high confinement into the guiding region. We consider a general inhomogeneous array of $N$ identical waveguides and continuous-wave propagating fields. The physical processes involved are described by the following system of equations \cite{Rai2012,Linares2008}
\begin{equation} \label{one}
\frac{d \hat{\mathcal{A}}_{j}}{d z}=\, i C_{0} (f_{j-1} \hat{\mathcal{A}}_{j-1}+ f_{j}\hat{\mathcal{A}}_{j+1}) +2 i \eta_{j} \hat{\mathcal{A}}_{j}^{\dag}, \\
\end{equation}
where $\hat{\mathcal{A}}_{0}=0$ and $\hat{\mathcal{A}}_{N+1}=0$, $f_{0}=f_{N}=0$ and $j=1,\dots, N$ is the individual mode index. $\hat{\mathcal{A}}_{j}\equiv \hat{\mathcal{A}}_{j}(z, \omega_{s})$ are monochromatic slowly-varying amplitude annihilation operators of signal (s) photons corresponding to the $j$th waveguide where $[\hat{\mathcal{A}}_{j}(z,\omega), \hat{\mathcal{A}}_{j'}^{\dag}(z,\omega')]=\delta(\omega-\omega')\delta_{j,j'}$. \Da{$\hat{\mathcal{A}}_{j}$ are thus the annihilation operators in the individual mode basis.} $\eta_{j}=g\, \alpha_{h,j} \equiv \vert\eta_{j}\vert\,e^{i\phi_{j}}$ is the effective nonlinear coupling constant corresponding to the $j$th waveguide, with $g$ the nonlinear constant proportional to $\chi^{(2)}$ and the spatial overlap of the signal and harmonic fields in each waveguide. $C_{j}=C_{0} f_{j}$ is the linear coupling constant between modes $j$ and $j+1$, and $z$ is the coordinate along the direction of propagation. Both the coupling and nonlinear constants depend on the signal frequency, $C_{0} \equiv C_{0}(\omega_{s})$ and $g \equiv g(\omega_{s})$. In the following, we set them as real without loss of generality. 

The ANW presents a large number of governing parameters that can be engineered \Da{to fit} a desired operation \Da{and application}. There are two types of parameters according to our ability to reconfigure them: i) the evanescent coupling profile $\vec{f}=(f_{1},\dots ,f_{N-1})$ \cite{Moison2009, Chapman2016, Weimann2016, Blanco2018}, the length of the sample $L$, the number of waveguides $N$ --and notably its parity--, and the wavevector and coupling periodical inversion periods ($\Lambda_{\Delta\beta}$, $\Lambda_{C}$) in quasi-phase matched structures \cite{Hum2007, Barral2019c} which can not be tuned once the sample is fabricated. In contrast with ii) the power and phase pump profile, given respectively by $\vec{\eta}=(\vert\eta_{1}\vert, \dots, \vert\eta_{N}\vert)$ and $\vec{\phi}=(\arg{(\eta_{1})}, \dots, \arg{(\eta_{N})})$, the coupling strength $C_{0}(\omega_{s})$ \cite{Iwanow2004} and the basis of detection \cite{Cai2017} which can be set according to a required operation and encoding of information. The large number of degrees of freedom available in a compact device is the main advantage of the ANW with respect to other approaches in the generation of multimode entangled states \cite{Su2013, Armstrong2015}. We demonstrate in this paper how to use these tunable parameters to generate CV multipartite entanglement and cluster states.

To that end, we first need to solve the propagation in the ANW. Equation (\ref{one}) can be solved numerically for a specific set of parameters $C_{j}$, $\eta_{j}$ and $N$, or even analytically if $N$ is small. However, it is difficult to gain physical insight from numerical or low-dimension analytical solutions due to the increasing complexity of the system with the number of waveguides. The problem of propagation in ANW is simplified by using the eigenmodes of the corresponding linear array of waveguides --the linear or propagation supermodes \cite{Kapon1984}. \Da{These linear supermodes are especially useful as an intermediate step to diagonalize the evolution of the system and they are used throughout section III.} These supermodes are coupled through the nonlinearity. We find a further simplification by using the eigenmodes of the full nonlinear system --the nonlinear or squeezing supermodes \cite{Patera2010}--, which are squeezed and fully decoupled but local ($z$-dependent). \Da{Thus, we use three complementary and useful bases in ANW: the individual mode ${\mathcal{A}}$, the linear supermode ${\mathcal{B}}$ and the nonlinear supermode ${\mathcal{C}}$ bases.} Quantum information can be naturally encoded in any of the three bases. A suitable implementation of the detection provides access to a given encoding \cite{Fabre2019}. \Final{Below we give general solutions to the propagation problem in the three bases, discuss the methods of detection suitable for each basis/encoding, present an example with an analytical solution and summarize the main features of the three encoding strategies.}

\section{Generation and detection of multimode squeezed states}\label{III}

\subsection{Propagation solution in the individual mode and supermode bases}
We derive below the general solutions to the propagation in ANW and present the different available detection methods depending on the encoding strategy. Considering coupling only between nearest-neighbour waveguides, a linear waveguide array (Equation (\ref{one}) with $\eta_{j}=0$) presents supermodes \Da{${\mathcal{B}}$},  i.e. propagation eigenmodes. In general, any linear waveguide array is represented by a Hermitian tridiagonal matrix --Jacobi matrix-- with non-negative entries and thus by a set of non-degenerate eigenvalues and eigenvectors given in terms of orthogonal polynomials \cite{Kapon1984, Efremidis2005, Bosse2017}. These eigenvectors (linear supermodes) form a basis and are represented by an orthogonal matrix $M\equiv M(\vec{f})$ with real elements $M_{k,j}$. The individual mode and linear supermode bases are related by \Da{$\hat{\mathcal{B}}_{k}=\sum_{j=1}^{N} M_{k,j}\,\hat{\mathcal{A}}_{j}$.}
The supermodes are orthonormal $\sum_{j=1}^{N}M_{k,j} M_{k',j}=\delta_{k,k'}$, and the spectrum of eigenvalues is \Da{$\lambda_{k}\equiv \lambda_{k}(C_{0}, \vec{f})$}. \David{We label the supermodes in a descending order with respect to their eigenvalue.} We focus on the relevant case of homogeneous coupling along propagation, i.e. $C_{j}$ does not depend on $z$. Using slowly-varying supermode amplitudes \Da{$\hat{{B}}_{k}=\hat{\mathcal{B}}_{k}\,e^{-i\lambda_{k}z}$} and the orthonormality property, the following equation for the linear propagation supermodes is obtained
\begin{equation}\label{Hei}
\frac{d\hat{{B}}_{k}}{d z}=\sum_{k'=1}^{N} \mathcal{L}_{k,k'} (z) \,\hat{{B}}_{k'}^{\dag}.
\end{equation}
The coupling matrix $\mathcal{L}(z)$ is the local joint-spatial supermode distribution of the ANW and its elements are given by
\begin{equation} \label{six}
\mathcal{L}_{k,k'} (z)=2i\sum_{j=1}^{N} \vert \eta_{j}\vert M_{k,j} M_{k',j}\,e^{i\{\phi_{j}-(\lambda_{k}+\lambda_{k'})z\}}.
\end{equation}
$\mathcal{L}(z)$ is a complex symmetric matrix which gathers the information about the spatial shape of the pump, i.e. amplitudes and phases in each waveguide, and the propagation supermodes coupling. 

\Final{Equation (\ref{Hei}) is ubiquitous in the context of multimode squeezing \cite{Fabre2019, Patera2010}. In other approaches the coupling matrix is however constant along propagation, i.e. $\mathcal{L}$ does not depend on $z$ \cite{Arzani2018}. $z$-dependence is a unique feature of the ANW. The general solution to Equation (\ref{Hei}) has thus to take into account space-ordering effects when necessary \cite{Barral2020A}. For instance, in single-pass $z$-dependent PDC, for gains generating squeezing above 12 dB \cite{Christ2013}. However, space-ordering effects can be neglected here since i) a low-gain regime --small $\vert \eta_{j} \vert$-- is crucial for individual-mode entanglement, indeed the light generated in each waveguide remains guided without coupling if $\vert \eta_{j} \vert > C_{0}$ \cite{Fiurasek2000}, ii) our starting point Equation (\ref{one}) is correct for small values of $C_{0}$ since next-to-nearest neighbor evanescent coupling should be included for large values of $C_{0}$ \cite{Kapon1984}, and iii) we study SPDC generated from vacuum where space-ordering corrections start at the third order --roughly as $\mathcal{O}(\vert \eta_{j} \vert^{3}) $\cite{Quesada2014}. The formal solution to Equation (\ref{Hei}) in this regime can thus be written as}
\begin{equation} \label{Bsol}
\begin{small}
\begin{pmatrix}
\vec{B}(z) \\
\vec{B}^{\dag}(z)
\end{pmatrix}= \exp \left\lbrace
\begin{pmatrix}
0 & \int_{0}^{z} \mathcal{L}(z') dz' \\
\int_{0}^{z} \mathcal{L}^{*}(z') dz' & 0
\end{pmatrix}
  \right\rbrace
  \begin{pmatrix}
\vec{B}(0) \\
\vec{B}^{\dag}(0)
\end{pmatrix},
\end{small}
\end{equation}
with $\vec{B}=(\hat{B}_{1}, \dots, \hat{B}_{N})^{T}$. The solution of Equation (\ref{Bsol}) can be simplified using the local nonlinear supermode basis $\hat{\mathcal{C}}$, given by $ \hat{\mathcal{C}}_{m}=\sum_{k=1}^{N} \Upsilon_{m,k}^{\dag}(z)\,\hat{{B}}_{k}$, where $\Upsilon(z)$ is an unitary matrix which diagonalizes the complex symmetric matrix $\int_{0}^{z} \mathcal{L}(z') dz'$ by a congruence transformation -- the Autonne-Takagi transformation --, such that $\Upsilon(z) \,[\int_{0}^{z} \mathcal{L}(z') dz' ] \,\Upsilon^{T}(z)= \Lambda(z)$, with $\Lambda(z)$ a real diagonal matrix with non-negative entries \cite{Cariolaro2016}. \David{Equation (\ref{Bsol}) in terms of nonlinear supermodes is thus simply given by}
\begin{align} \label{nineC}
\hat{\mathcal{C}}_{m}(z)=\cosh[{r}_{m}(z)] \,\hat{\mathcal{C}}_{m}(0)+\sinh[{r}_{m}(z)]\,\hat{\mathcal{C}}_{m}^{\dag}(0),
\end{align}
where ${r}_{m}(z)=\Lambda_{m,m}(z)$ are the downconversion gains at a propagation distance $z$. Each local nonlinear supermode thus appears as a single-mode squeezed state. The $z$-dependence of the nonlinear supermodes is a trademark of the ANW. The evanescent coupling indeed produces a phase mismatch between the pump and the generated signal waves which results in a $z$-dependent interaction. This coupling-based phase mismatch affects the amount of squeezing and entanglement generated in the ANW. The relation between the nonlinear supermodes and the individual modes is the following
\begin{equation}\label{NSup}
 \hat{\mathcal{C}}_{m}=\sum_{k=1}^{N} \sum_{j=1}^{N} (\Upsilon_{m,k}^{\dag}(z)\, M_{k,j}\,e^{-i\lambda_{k} z}) \hat{\mathcal{A}}_{j} .
 \end{equation}
This expression encapsulates the mechanisms at play in the ANW: the evanescent coupling generates the linear supermodes ($M_{k,j}$) which get a phase due to propagation ($\lambda_{k} z$) and the nonlinearity couples them locally ($\Upsilon_{m,k}^{\dag}(z)$). In terms of the individual modes, the solution to the nonlinear system is thus
\begin{align}\nonumber
\hat{\mathcal{A}}_{j}(z)=&\sum_{k,m,j'=1}^{N} (M_{j,k} \Upsilon_{k,m}(z) M_{m,j'}\, e^{i \lambda_{k} z}) \\ \label{IndGenSol}
&\{\cosh[{r}_{m}(z)]\,\hat{\mathcal{A}}_{j'}(0)+\sinh[{r}_{m}(z)] \,\hat{\mathcal{A}}_{j'}^{\dag}(0)\}.
\end{align}
Equations (\ref{Bsol}), (\ref{nineC}) and (\ref{IndGenSol}) are the general solutions to the propagation problem in ANW in the linear supermodes, nonlinear supermodes and individual mode bases, respectively. These three solutions represent a resource for encoding quantum information. Our results generalize those previously obtained in driven quantum walks in ANW \cite{Kruse2013, Solntsev2014, Hamilton2014}. 

Since we are interested in CV squeezing and entanglement, we will also use along the paper the field quadratures  $\hat{x}_{j}$, $\hat{y}_{j}$, where $\hat{x}_{j}=(\hat{\mathcal{A}}_{j}+\hat{\mathcal{A}}_{j}^{\dag})$ and $\hat{y}_{j}=i (\hat{\mathcal{A}}_{j}^{\dag}-\hat{\mathcal{A}}_{j})$ are, respectively, the amplitude and phase quadratures corresponding to a signal optical mode \Da{in the individual mode basis $\mathcal{A}$. Field quadratures for the linear $\mathcal{B}$ and nonlinear $\mathcal{C}$ supermodes are defined in a similar way}. \Final{In the basis of individual modes, with quadratures $\hat{\xi}=(\hat{x}_{1},\dots, \hat{x}_{N}, \hat{y}_{1},\dots,\hat{y}_{N})^T$, the general solution of Equation (\ref{IndGenSol}) can be written as $\hat{\xi}(z)={S}(z)\, \hat{\xi}(0)$, with ${S}(z)$ a symplectic matrix which contains all the information about the propagation of the quantum state of the system. The quantum states generated in ANW are Gaussian.} The most interesting observables in Gaussian CV are the second-order moments of the quadrature operators, properly arranged in the covariance matrix ${V}$ \cite{Adesso2014}. For a quantum state initially in vacuum, the covariance matrix at any plane $z$ is given by ${V}(z)={S}(z)\,{S}^{T}(z)$, with 1 the value of the shot noise related to each quadrature in our notation. Evolution of variances $V(\xi_{i}, \xi_{i})$ and quantum correlations $V(\xi_{i}, \xi_{j})$ can be obtained at any length from the elements of this matrix. The covariance matrix can be written in the following way
\begin{equation}\label{BMD}
V(z)={R}_{1}(z) K^{2}(z) {R}^{T}_{1}(z),
\end{equation} 
where we have applied a Bloch-Messiah decomposition of ${S}(z)$ \cite{Braunstein2005a}. $K^{2}(z)\equiv K^{2}(\vec{\eta}, \vec{\phi}, z)=\diag\{e^{2 r_{1}(z)}, \dots, e^{2 r_{N}(z)}, e^{-2 r_{1}(z)}, \dots, e^{-2 r_{N}(z)}\}$ is the covariance matrix in the nonlinear supermode basis and $R_{1}(z)\equiv R_{1}(\vec{\eta}, \vec{\phi}, z)$ the symplectic transformation matrix between the individual and nonlinear supermode basis, equivalent to Equation (\ref{NSup}) for complex fields. In fact, the Bloch-Messiah and the Autonne-Takagi approaches are fully equivalent \cite{Cariolaro2016}. The $m$th nonlinear supermode is squeezed and thus nonclassical if $K^{2}_{N+m}(z)= e^{-2 r_{m}(z)}<1$.

\subsection{\Da{Detection schemes}}\label{IIIb}

The measurement of quantum noise variances and correlations is carried out by multimode balanced homodyne detection (BHD) and access to the quantum information encoded in the individual or any of the supermode bases depends on a suitable BHD realization \cite{Fabre2019}. Our photonic entanglement synthesizer is certainly compatible with a fully fibered implementation of detection via off-the-shelf telecom components \Final{as shown in Figure \ref{F1} \cite{Kaiser2016}.} 

The three different bases where quantum information can be naturally encoded in the ANW correspond to different choice and engineering of LOs in multimode BHD: \Final{i) In the individual mode basis each LO is independent. Only two detectors are necessary in order to completely characterize any multimode quantum state. The variance measured in each mode and the quantum correlations between any pair of modes allow to reconstruct the full covariance matrix associated to the generated quantum state. Beyond quantum state tomography, to perform a quantum information protocol, it is necessary to use the same number of LOs and detectors in the multimode BHD as the number of modes involved in the protocol. ii) The detection in an arbitrary supermode basis is based on the combination of the SPDC output fields with a spatially reconfigurable multimode local oscillator (LO shaper in Figure $\ref{F1}$), a spatial analogue to what has been pioneered in the spectral domain in \cite{Cai2017}. The LO shaper sets the relative amplitudes and phases of the multimode LO to select a given supermode, and a controllable global phase ($\theta_{0}$ in Figure $\ref{F1}$) selects the quadrature to be measured \cite{Raymer1996, Dauria2009}. This shaped LO can indeed be set successively as the elements of a supermode basis enabling the measurement of the full covariance matrix in that basis. This measurement method can be implemented in fibered or in bulk optics through respectively fibered multimode BHD or bulk-optics single-mode BHD with a spatially shaped LO.} \Da{We outline} that every pump configuration and length of the sample $L$ produces a different nonlinear supermode basis as stated by Equation (\ref{NSup}). Thus, the detection basis has to be reconfigured for every pump distribution when this encoding strategy is selected. Our theoretical model is hence crucial to decide \Da{operation and fabrication} parameters of the ANW so that it provides a sought-after operation. 

\subsection{\Da{Example: squeezing in the three mode bases for a flat pump profile}}\label{IIIc}

\begin{figure}[t]
  \centering
    \subfigure{\includegraphics[width=0.48\textwidth]{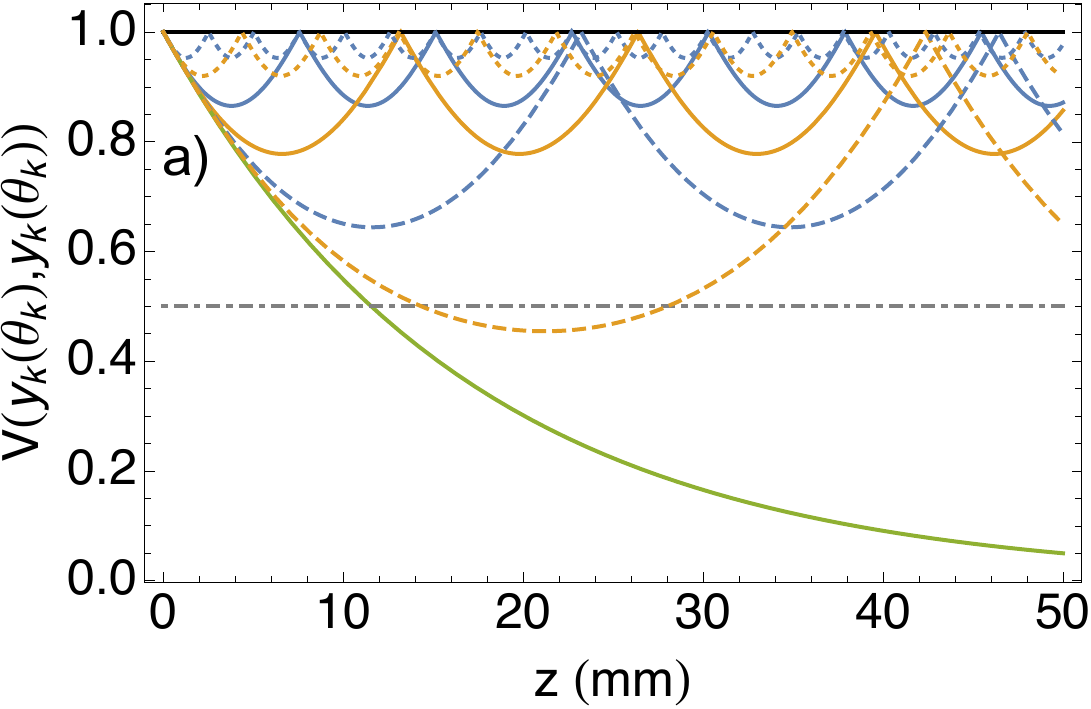}}
    \subfigure{\includegraphics[width=0.48\textwidth]{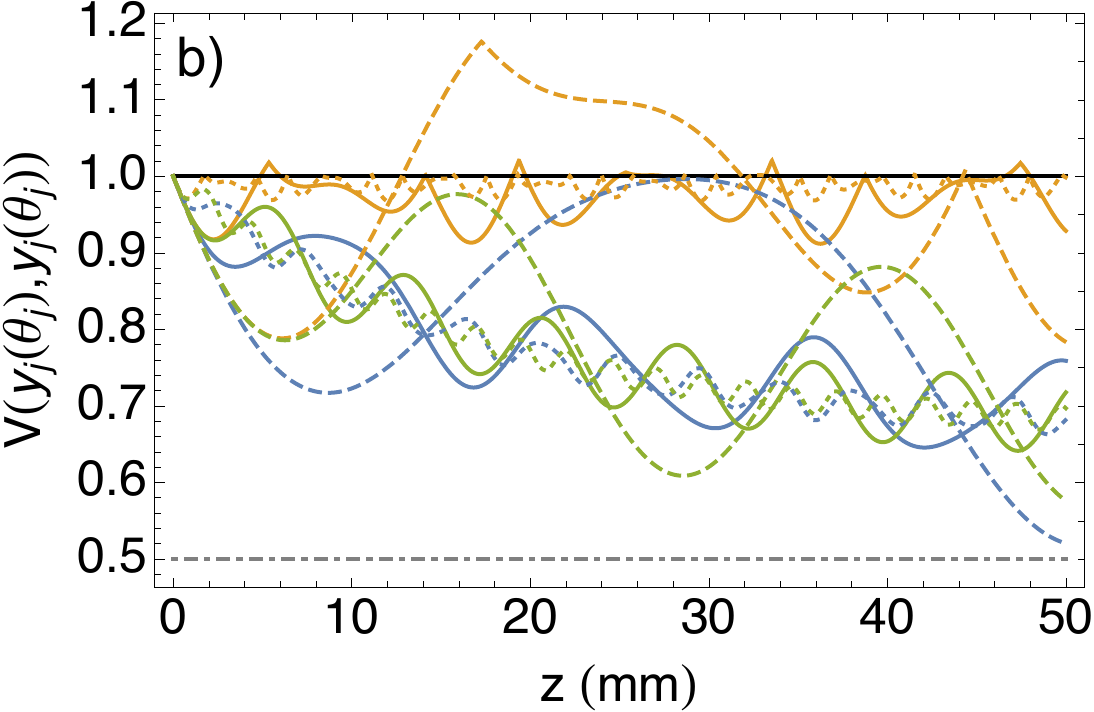}}
\vspace {0cm}\,
\hspace{0cm}\caption{\label{F2}\small{\Da{Evolution of squeezing in a five-waveguides homogeneous coupling profile nonlinear array pumping with a flat pump profile. a) Maximum linear supermode squeezing $[V(y_{k}(\theta_{k}), y(\theta_{k}))<1]$. Equivalent curves are retrieved for nonlinear supermode squeezing. The $k=3$ supermode is efficiently squeezed independently of the coupling strength (solid, green). The other supermodes are degenerate two by two: $k=1,5$ (blue) and $k= 2,4$ (orange). b) Maximum individual mode squeezing $[V(y_{j}(\theta_{j}),y_{j}(\theta_{j}))<1]$. $j=1,5$ (blue), $j= 2,4$ (orange) and $j=3$ (green). $C_{0}=0.08, 0.24$ and $0.72$ mm$^{-1}$ are dashed, solid and dotted, respectively. $\eta=0.015$ mm$^{-1}$. 3 dB squeezing level in dot-dashed, gray. }}}
\end{figure}

\Final{Suitable manipulation of individual power and phase pump fields by means of off-the-shelf elements as fiber attenuators and phase shifters, and V-groove arrays linking the fiber-optics elements to the ANW enable an on-demand pump distribution engineering (Figure \ref{F1}). The pump profile couples the propagation supermodes generating the joint-spatial supermode distribution given by Equation (\ref{six}). In general, this generates complicated connections between the linear supermodes. However, the orthogonality and symmetry properties of the linear supermodes lead to analytical solutions in some cases. An outstanding simplification of the system is obtained when pumping all the waveguides with the same power and phase, which we refer to as the flat pump profile. We show below the features of this configuration in the three bases:

\paragraph{Linear supermode basis.} When all waveguides are equally pumped such that $\eta_{j}=\eta$, the Equation (\ref{six}) is notably simplified to $\mathcal{L}_{k,k'} (z)=2i  \,\eta\, \delta_{k,k'} \exp\{-i (\lambda_{k}+\lambda_{k'})z\}$, where we have used the orthonormality of the linear supermodes. This pump configuration diagonalizes the joint-spatial distribution making every linear supermode independently squeezed. The solution of Equation (\ref{Hei}) in terms of linear supermodes $\mathcal{B}_{k}$ can then be written as
\begin{equation} \label{k-sol}
\hat{\mathcal{B}}_{k}(z)=\cos(F_{k} z)\hat{\mathcal{B}}_{k}(0)+i\frac{\sin(F_{k} z)}{F_{k}}  [\lambda_{k}\hat{\mathcal{B}}_{k}(0)+2\eta\,\hat{\mathcal{B}}_{k}^{\dag}(0)],
\end{equation}
with $F_{k}=\sqrt{\lambda_{k}^{2}-4\vert\eta\vert^{2}}$. This solution is exact for any gain regime \cite{Barral2020A}. It generalizes the solution found for the nonlinear directional coupler ($N=2$) in ref. \cite{Barral2017} and remains valid for any number of waveguides $N$, evanescent coupling profile $\vec{f}$, and propagation distance $z$. Note that Equation (\ref{k-sol}) for $\vert \lambda_{k} \vert > 2 \vert \eta \vert $ and $\vert \lambda_{k} \vert  < 2 \vert \eta \vert $ are, respectively, the solutions of a non-phase-matched and phase-matched degenerate parametric amplifiers \cite{Mollow1967}. For typical evanescent coupling, nonlinearities and pump powers found in quadratic ANW, $F_{k}$ is real for the majority of supermodes. In this case, the $k$th propagation supermode periodically oscillates between a maximum and zero with oscillation periods $L_{k}=\pi/(2 F_{k})$. The period $L_{k}$ depends on the pump power and, markedly, on the evanescent coupling profile $\vec{f}$ through $\lambda_{k}$. Interestingly, waveguide arrays with odd number of identical waveguides $N$ exhibit a propagation eigenmode with zero eigenvalue $\lambda_{(N+1)/2}=0$ \cite{Efremidis2005}. This supermode is phase-matched all along the propagation and its squeezing is maximum \cite{Barral2020L}. 

We exhibit the features discussed above with a specific example in an array with $N=5$ waveguides. Figure \ref{F2}a shows the evolution of maximum linear supermode squeezing [$V(y_{k}(\theta_{k}), y(\theta_{k}))<1$] over a generalized phase quadrature ${y}_{k}(\theta_{k})$ \cite{Note00}. We set the $k$th LO phase as $\theta_{k}(z)$ maximizing the $k$th linear supermode squeezing \cite{Barral2020A}. We set $\vert \eta \vert=0.015$ mm$^{-1}$, a realistic value comparable with that obtained in ref. \cite{Lenzini2018}. We choose a homogeneous coupling profile $f_{j}=1$. The propagation constants related to each propagation supermode with this coupling profile are $\lambda_{k}=2 C_{0} \cos[k \pi / (N+1)]$ \cite{Kapon1984}. We show the effect of the coupling strength $C_{0}$ on the squeezing with three different cases: 0.08 mm$^{-1}$ (dashed), 0.24 mm$^{-1}$ (solid), and 0.72 mm$^{-1}$ (dotted). The $k=3$ supermode is the only nondegenerate one (green). It is independent of the value of $C_{0}$, builds up continuously and is efficiently squeezed. The amount of squeezing is degenerate two by two for the other supermodes. The maximum squeezing in these supermodes ($k=1,5$ in blue, $k=2,4$ in orange) decreases as the coupling strength $C_{0}$ increases. 

The properties of the linear supermodes thus give insight about the properties of the squeezed light generated in the array in this pump configuration. We have found and analyzed similar analytical solutions for other pump configurations in this basis \cite{Barral2020A}. 

\paragraph{Nonlinear supermode basis.} The flat pump configuration diagonalizes directly the system in the linear supermode basis. After \Final{an appropriate} phase-space rotation, carried out by the LO in the multimode BHD, the system is fully diagonal in the phase space and thus linear and nonlinear supermodes are degenerate. Note that for the example presented above, the nonlinear supermodes obtained via Autonne-Takagi (or Bloch-Messiah) diagonalization exhibit the same levels of squeezing $K^{2}_{N+m}(z)$ (not shown) as those shown in Figure \ref{F2}a for linear supermodes, but a different spatial profile at each propagation length $z$ \cite{Barral2020A}.

\paragraph{Individual mode basis.} The solution in the individual mode basis is straightforwardly obtained from Equation (\ref{k-sol}) with the basis change $\hat{\mathcal{A}}_{j}=\sum_{k=1}^{N} M_{k,j}\,\hat{\mathcal{B}}_{k}$. Note that this is analogous to input a set of $N$ independently squeezed states in an $N\times N$ interferometer given by the orthogonal matrix $M\equiv M(\vec{f})$. This configuration can thus produce entanglement in the individual mode basis as we will show in the next section.

Figure \ref{F2}b shows the evolution of maximum individual mode $\mathcal{A}_{j}$ squeezing $[V(y_{j}(\theta_{j}),y_{j}(\theta_{j}))<1]$ over a generalized phase quadrature ${y}_{j}(\theta_{j})$ \cite{Note00} for the same configuration and coupling strengths than Figure \ref{F2}a. In this case we set independently the LO phase of each BHD as $\theta_{j}(z)$ to maximize the squeezing of the $j$th individual mode. Note that there are distances where the individual mode squeezing for modes $j=2,4$ (orange) is over the shot noise. In contrast to the linear supermode case, the evolution of squeezing is complex in the individual mode basis and it is difficult to obtain physical insight directly from it. Moreover, this configuration presents quantum correlations (not shown) that can lead to entanglement.}

The possibilities of our concept go nonetheless considerably beyond the flat pump profile. The ANW presents a large number of degrees of freedom that can be suitably tuned in order to yield a desired mode of operation. We tackle this question in depth and exhibit the versatility of the ANW addressing a crucial and fundamental issue in quantum information processing: the generation of multipartite entanglement. \Final{We showcase two remarkable axes in the most appropriate encoding: i) in section \ref{IV}, partial numerical optimization on the pump phase profile in the individual mode basis, and ii) in section \ref{V}, full numerical optimization on all available parameters to demonstrate versatility in the individual mode basis and in optimized nonlocal bases built from nonlinear supermodes.} 

\subsection{\Da{Encoding strategies summary}}\label{IIId}

\Da{To end this section, we sketch an informal practical landscape of the introduced bases in the ANW: i) the elements of the individual mode basis $\mathcal{A}$ are \Final{modes associated with a single waveguide, they can be distributed to spatially distant locations. All elements of this basis can be jointly measured} and the detection basis is fixed. Squeezing is not optimal in this basis. It is directly connected to applications in quantum networks and can be used in existing communication, sensing and computing protocols (see section \ref{VI} and especially table \ref{Table2}); ii) the elements of the linear supermode basis $\mathcal{B}$ are collective modes. Only one mode can be measured at a time and the detection basis is fixed (pump profile and device length independent). Squeezing is in general not optimal. $\mathcal{B}$ is useful to derive trends analytically \MOD{\cite{Barral2020A}} as in Figure \ref{F2}a that can be applied to the other encodings (${\mathcal{A}}$, ${\mathcal{C}}$) and this basis is well suited to quantum simulation; iii) the elements of the nonlinear supermode basis ${\mathcal{C}}$ are collective modes. Only one mode can be measured at a time and the detection basis is variable (pump profile and device length dependent). Squeezing is optimal in this basis. Entanglement is engineered in the measurement stage. It is close to a rich niche in the spectral domain \cite{Armstrong2012,Roslund2013,Cai2017} and is well suited to \Final{engineer quantum states and study their features}.}

\section{Multipartite entanglement of individual modes in ANW}   \label{IV}

{Multipartite entangled states are a key resource for quantum computing, through the measurement based framework \cite{Raussendorf2001, Menicucci2006}, and for quantum key distribution networks, obtaining higher secret key rates with respect to bipartite entangled states \MOD{and extending the range of application through all-photonic quantum repeaters \cite{Epping2017, Hasegawa2019}}. Measuring multipartite full inseparability in CV systems requires the simultaneous fulfillment of a set of conditions which leads to genuine multipartite entanglement when pure states are involved \cite{vanLoock2003}. This criterion, known as van Loock - Furusawa inequalities (VLF), can be easily calculated from the elements of the covariance matrix $V$ in a given basis. \David{Full $N$-partite inseparability is guaranteed if the following $N-1$ inequalities are simultaneously violated in a basis with modes labelled by $i$ \cite{vanLoock2003}}
\Da{\begin{align} \nonumber
&\rho_{i}\equiv V[{x}_{i}({\theta_{i}}) - {x}_{i+1}(\theta_{i+1})] + \\   \label{VLF}
&V[{y}_{i}({\theta_{i}}) + {y}_{i+1}({\theta_{i+1}}) +\sum_{i'\neq i,i+1}^{N} G_{i'} \,{y}_{i'}(\theta_{i'})] \geq 4,
\end{align}
where are ${x}_{i}({\theta_{i}})$ and ${y}_{i}({\theta_{i}})$ are generalized quadratures} \cite{Note00}, $\vec{\theta}\equiv (\theta_{1}, \dots, \theta_{N})$ is the measurement LO phase profile, $\vec{G}\equiv (G_{1}, \dots, G_{N})$ are $N$ real parameters which are set by optimization --a BHD gain profile-- \cite{Note0}.

\Da{In general, we can prepare the state in such a way that the generated SPDC light is multipartite entangled. In other words,} we find that there is a set of pump amplitude $\vec{\eta}$ and phase $\vec{\phi}$ profiles, as well as LO phase $\vec{\theta}$ and BHD gain $\vec{G}$ profiles, which minimize the value of Equations (\ref{VLF}) for a given set of fixed parameters of the array --coupling profile $\vec{f}$, number of waveguides $N$, and length of the sample $L$-- and choice of encoding. Below we demonstrate that the ANW is a versatile source of multipartite entanglement through optimization. We choose the individual mode basis as measurement basis \Da{as it is the most accessible basis and entanglement can be distributed in a network}, and compare the entanglement among these modes produced by the flat pump profile introduced in Section \ref{IV}, where only the BHD parameters $\{\vec{G}, \vec{\theta}\}$ can be tuned, with the entanglement yielded if the pump phase profile is also tuned $\{\vec{\phi}, \vec{G}, \vec{\theta}\}$ in an optimization procedure. \Final{Note that the elements of the linear supermode basis can be also entangled. By contrast, the elements of the nonlinear supermode basis are not entangled -- i.e. are independent--, but they can be used as a starting point to engineer entangled states through suitable transformations. More details are presented in sections \ref{V} and \ref{VI}.}

Firstly, as an example of a non-optimal procedure we focus on the cases shown in Figure \ref{F3}: $N=5$ ANW with a homogeneous coupling profile $f_{j}=1$ and a flat pump power distribution ($\vert\eta_{j}\vert=\vert\eta\vert$, $\Delta\phi=\phi_{j+1}-\phi_{j}=0$). We analyze the impact of the coupling strength $C_{0}$ on the multipartite entanglement as we noticed in \Da{Figures \ref{F2}a and \ref{F2}b} it has an impact on squeezing. \David{In this case we set the individual mode basis and adapt the detection}, i.e. the BHD parameters: the LO phases $\vec{\theta}$ and the electronic gains $\vec{G}$. We use the sum of the four inequalities $F_{M}(\vec{G}, \vec{\theta})=\sum_{j=1}^{4} \rho_{j}$ as the fitness function to optimize. We use an evolution-strategy algorithm to tackle the optimization problems found along the paper \cite{Beyer2002}. Our optimization algorithm adjusts 10 parameters to find the minimum of $F_{M}$. Figure \ref{F3} shows two by two degenerate inequalities found for three different values of coupling strength $C_{0}$: 0.08 mm$^{-1}$ (dashed), 0.24 mm$^{-1}$ (solid), and 0.72 mm$^{-1}$ (dotted). The three cases present regions where fully multipartite entanglement is achieved ($\rho_{j}<4$). The lower the value of $C_{0}$, the larger the violation of the inequalities as expected from \Da{Figure $\ref{F2}$a}. 
\begin{figure}[t]
  \centering
    {\includegraphics[width=0.48\textwidth]{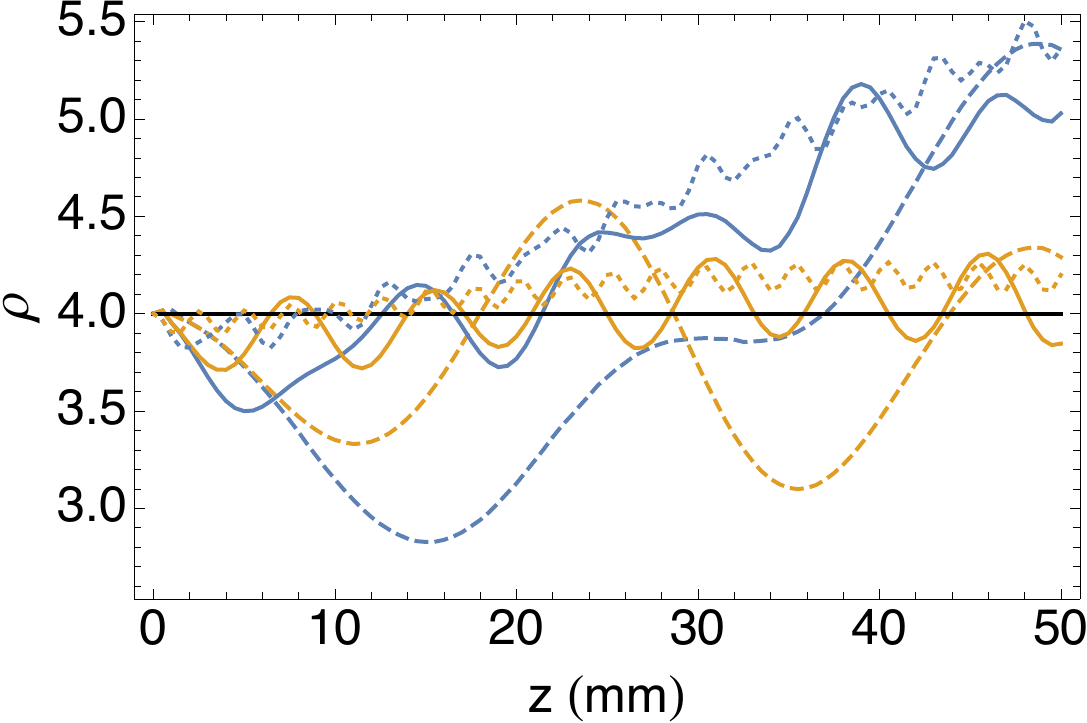}}
    \vspace{0cm}
\hspace{0cm}\caption{\label{F3}\small{Multipartite entanglement in the individual mode basis versus propagation for a flat pump profile ($\vert\eta_{j}\vert=\vert\eta\vert$, $\Delta\phi=0$) in a 5-waveguides ANW. \David{Van Loock - Furusawa inequalities obtained optimizing the BHD parameters $\{\vec{G}, \vec{\theta}\}$}. Simultaneous values under the threshold value $\rho=4$ (black) imply CV pentapartite entanglement (four inequalities, degenerate two by two: blue and orange curves). $C_{0}=0.08, 0.24$ and $0.72$ mm$^{-1}$ are dashed, solid and dotted, respectively. $\eta=0.015$ mm$^{-1}$.}}
\end{figure} 
\begin{figure}[t]
  \centering
   \subfigure{\includegraphics[width=0.48\textwidth]{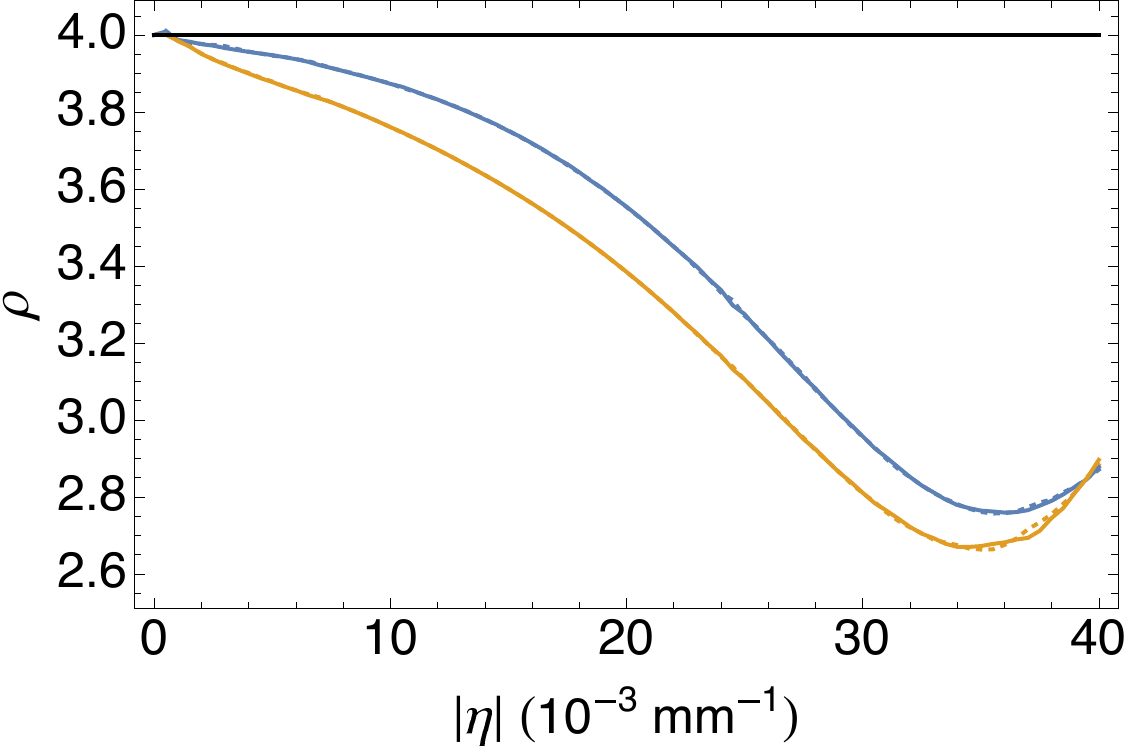}}    
  \vspace {0cm}\,
\hspace{0cm}\caption{\label{F4}\small{Multipartite entanglement in the individual mode basis versus flat pump-power profile ($\vert\eta_{j}\vert=\vert\eta\vert$, $\Delta\phi \neq 0$) in a 5-waveguides ANW. \David{Van Loock - Furusawa inequalities obtained optimizing the pump phases and BHD parameters $\{\vec{\phi}, \vec{G}, \vec{\theta}\}$}. Simultaneous values under the threshold value $\rho=4$ (black) imply CV pentapartite entanglement (four inequalities, N=5). j=1 (dotted blue), j=2 (dotted orange), j=3 (solid orange) and j=4 (solid blue). $C_{0}=0.24$ mm$^{-1}$. $z=30$ mm.}}
\end{figure}


Let us move now to further optimized example. We study the same ANW at a fixed length $z=30$ mm and coupling strength $C_{0}=0.24$ mm$^{-1}$ where we can additionally tune the individual pump phases ($\vert\eta_{j}\vert=\vert\eta\vert$, $\Delta\phi \neq 0$). Note that there is no entanglement at this distance for $\Delta\phi=0$ as shown in Figure \ref{F3} ($\rho_{1,4}=4.51$ solid blue, $\rho_{2,3}=4.23$ solid orange). We use again the sum of the four inequalities $F_{M}(\vec{\phi}, \vec{G}, \vec{\theta})$ as the fitness function to optimize, now with 4 extra parameters related to the relative pump phases. Figure \ref{F4} shows the four inequalities for five propagating modes for different values of power per waveguide $\vert\eta\vert$, among them $\vert\eta\vert=0.015$ mm$^{-1}$, the case shown in Figure \ref{F3}. Genuine multipartite entanglement is obtained for any value of $\vert\eta\vert$ shown. Remarkably, the simultaneous violation of the four inequalities at $\vert\eta\vert=0.015$ mm$^{-1}$ by only optimizing the pump phases exhibits the versatility of our approach. Note that the above optimization procedure represents a lower bound on the violations based on the fitness function we have chosen and our optimization algorithm. Hence, there can be other sets of parameters which present larger violations of the VLF inequalities.

In the next section we broaden the spectrum of applications of the ANW to the optimized generation of multipartite states with a specific geometry: the cluster states. The interest in these states lies in that they are the essential resource of MBQC and one of the main actors in the quest for a photonic quantum computer \cite{Gu2009}. 


\section{Optimization for cluster states} \label{V}

MBQC relies on the availability of a large multimode entangled state on which a specific sequence of measurements is performed. The choice of natural or exotic bases widens the range of application in MBQC \cite{Ferrini2013}. In our case the cluster states can be encoded in the individual mode basis or any other basis of the array.

An ideal CV cluster state is a simultaneous eigenstate of specific quadrature combinations called nullifiers \cite{Raussendorf2001, Menicucci2006}. Cluster states are associated with a graph or \MOD{adjacency matrix $J$ reflecting those nullifiers. The nodes of the graph represent the modes of the cluster state in a given basis, and the edges the entanglement connections among the modes. Moreover, the labelling of the nodes can be suitably set to maximize the entanglement. The nullifiers are given by
\begin{equation}\label{Nul}
\hat{\delta}_{i}\equiv \hat{x}_{i}(\theta_{i}+\pi/2)-\sum_{i'=1}^{N} J_{i, i'}\, \hat{x}_{i'}(\theta_{i'}) \quad \forall i=1,\dots, N,
\end{equation}
where $J$ is the adjacency matrix and $ \hat{x}_{i}(\theta_{i})$ is the $i$th generalized quadrature in a given basis \cite{Note00}. We consider unit-weight cluster states with $J_{i,i'} = 1$ for modes $i$ and $i'$ being nearest neighbors in the graph and all the other entries of $J$ are zero.} \NewText{The variances} of the nullifiers tend to zero in the limit of infinite squeezing. Experimentally, a cluster state can be certified if two conditions are satisfied: i) the noise of a set of normalized nullifiers lies below shot noise
\begin{equation}\label{VNul}
V(\bar{\delta}_{i})<1 \quad \forall i=1,\dots, N,
\end{equation}
where $\bar{\delta}_{i}\equiv \delta_{i}/\sqrt{1+n(i)}$ is the normalized nullifier and $n(i)$ is the number of nearest neighbours to the $i$th node of the cluster, and ii) the cluster state is fully inseparable, i.e. it violates a set of VLF inequalities \cite{vanLoock2003, Yukawa2008}.

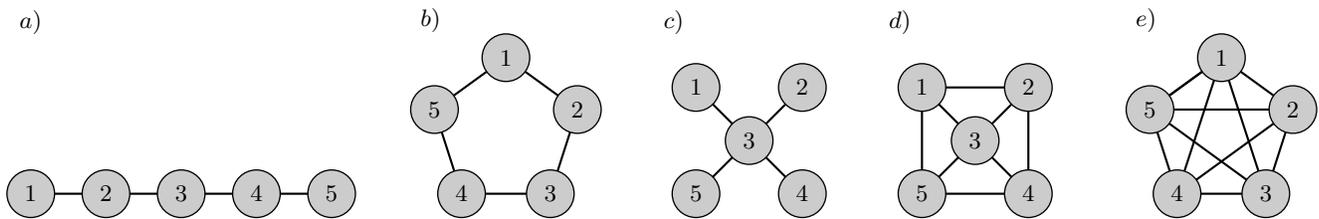
\begin{figure*}[t]
\begin{center}
\begin{tikzpicture}
\node at (0,1.57+1.414) {$a)$};
  \GraphInit[vstyle=Dijkstra]
  \SetVertexNormal[Shape=circle,FillColor=black!20]
  \Vertex[x=0,y=0.707,L=$1$]{A}    
  \Vertex[x=1,y=0.707,L=$2$]{B}    
  \Vertex[x=2,y=0.707,L=$3$]{C}
  \Vertex[x=3,y=0.707,L=$4$]{D}
  \Vertex[x=4,y=0.707,L=$5$]{E}
  \tikzset{EdgeStyle/.style={-}}
  \Edge (A)(B)
  \Edge (B)(C)
  \Edge (C)(D)
  \Edge (D)(E)
\end{tikzpicture}
\hspace{0.5cm}
\begin{tikzpicture}
\node at (-1,0.5) {$b)$};
  \GraphInit[vstyle=Dijkstra]
  \SetVertexNormal[Shape=circle,FillColor=black!20]
  \Vertex[L=$1$]{A}    
  \Vertex[x=0.95,y=-0.69,L=$2$]{B}    
  \Vertex[x=0.59,y=-1.81,L=$3$]{C}
  \Vertex[x=-0.59,y=-1.81,L=$4$]{D}
  \Vertex[x=-0.95,y=-0.69,L=$5$]{E}
  \tikzset{EdgeStyle/.style={-}}
  \Edge (A)(B)
  \Edge (B)(C)
  \Edge (C)(D)
  \Edge (D)(E)
  \Edge (E)(A)
\end{tikzpicture}
\hspace{0.5cm}
\begin{tikzpicture}
\node at (-1,1.57) {$c)$};
  \GraphInit[vstyle=Dijkstra]
  \SetVertexNormal[Shape=circle,FillColor=black!20]
  \Vertex[L=$3$]{A}    
  \Vertex[x=0.707,y=0.707,L=$2$]{B}    
  \Vertex[x=0.707,y=-0.707,L=$4$]{C}
  \Vertex[x=-0.707,y=-0.707,L=$5$]{D}
  \Vertex[x=-0.707,y=0.707,L=$1$]{E}
  \tikzset{EdgeStyle/.style={-}}
  \Edge (A)(B)
  \Edge (A)(C)
  \Edge (A)(D)
  \Edge (A)(E)
\end{tikzpicture}
\hspace{0.5cm}
\begin{tikzpicture}
\node at (-1,1.57) {$d)$};
  \GraphInit[vstyle=Dijkstra]
  \SetVertexNormal[Shape=circle,FillColor=black!20]
  \Vertex[L=$3$]{A}    
  \Vertex[x=0.707,y=0.707,L=$2$]{B}    
  \Vertex[x=0.707,y=-0.707,L=$4$]{C}
  \Vertex[x=-0.707,y=-0.707,L=$5$]{D}
  \Vertex[x=-0.707,y=0.707,L=$1$]{E}
  \tikzset{EdgeStyle/.style={-}}
  \Edge (A)(B)
  \Edge (A)(C)
  \Edge (A)(D)
  \Edge (A)(E)
  \Edge (B)(C)
  \Edge (C)(D)
  \Edge (D)(E)
  \Edge (E)(B)
\end{tikzpicture}
\hspace{0.5cm}
\begin{tikzpicture}
\hspace{0.25cm}
\node at (-1,0.5) {$e)$};
  \GraphInit[vstyle=Dijkstra]
  \SetVertexNormal[Shape=circle,FillColor=black!20]
  \Vertex[L=$1$]{A}    
  \Vertex[x=0.95,y=-0.69,L=$2$]{B}    
  \Vertex[x=0.59,y=-1.81,L=$3$]{C}
  \Vertex[x=-0.59,y=-1.81,L=$4$]{D}
  \Vertex[x=-0.95,y=-0.69,L=$5$]{E}
  \tikzset{EdgeStyle/.style={-}}
  \Edge (A)(B)
  \Edge (B)(C)
  \Edge (C)(D)
  \Edge (D)(E)
  \Edge (E)(A)
    \Edge (A)(C)
      \Edge (A)(D)
        \Edge (A)(E)
          \Edge (B)(D)
          \Edge (B)(E)
          \Edge (C)(E)
\end{tikzpicture}
\end{center}
\caption{\label{F9}\small{Some 5-nodes graphs generated in the ANW. a) Linear, b) Pentagon, c) Star, d) Square pyramid, and e) Maximally connected pentagon --or GHZ--.}}
\end{figure*}
\begin{table*}[]
\centering
\begin{tabular}{c c c c c c}
\hline\hline
Graph    &   Nullifiers $\{V(\bar{\delta}_{i})\}$  & $\vec{\eta} \times 10^{2}$ mm$^{-1}$ & $\vec{\phi}/ \pi$ & $\vec{\theta}/\pi$   \\[0.5ex]
\hline
Linear & $\{0.20, 0.39, 0.37, 0.38, 0.20\}$ &$\{9.2, 8.9, 9.1, 9.1, 9.2\}$ & $-0.50 \times \{1, 1, 1, 1, 1\}$ & $\{0,0,0,0,0\}$ \\[1ex]
Pentagon & $\{0.59, 0.73, 0.09, 0.34, 0.11\}$ &\{8.7, 4.9, 3.4, 1.9, 8.7\} & \{1.59, 0.87, 1.06, 1.34, 0.63\} &\{0.60, 0.19, -1.00, -0.88, 0.20\}\\[1ex]
Star &$\{0.40, 0.41, 0.54, 0.41, 0.40\}$  & \{6.1, 3.4, 9.5, 3.4, 6.1\} & \{1.02, 0.02, 0.02, 0.01, -0.98\} & \{-0.24, 0.26, 0.26, 0.26, -0.24\}  \\ [1ex] 
Pyramid &$\{0.33, 0.12, 0.57, 0.18, 0.19\}$ &$\{4.3, 9.3, 0.0, 7.3, 3.0\}$ &$\{-0.11, 0.22, 0.65, 1.29, 1.00\}$ & $\{0.05, 0.27, 0.40, -0.80, 0.00\}$\\ [1ex]
GHZ &$\{0.40, 0.41, 0.54, 0.41, 0.40\}$  & \{6.1, 3.4, 9.5, 3.4, 6.1\} & \{1.02, 0.02, 0.02, 0.01, -0.98\} & \{0.26, 0.76, 0.26, 0.76, 0.26\}  \\ [1ex] 
\hline
\end{tabular}
\caption{\label{Table1}\small{Individual mode basis cluster state generation in a 5-waveguides ANW with homogeneous coupling profile for linear, pentagon, star, square pyramid and GHZ graphs. We show the value of the nullifiers $\{V(\bar{\delta}_{i})\}$, pump power profile $\vec{\eta}$, pump phase profile $\vec{\phi}$ and local oscillator phase profile $\vec{\theta}$. Simultaneous values of $V(\bar{\delta}_{i})$ under the shot noise threshold $V(\bar{\delta}_{i})=1$ are a signature of \NewText{successful} cluster generation. $C_{0}=0.24$ mm$^{-1}$. $z=30$ mm.}}
\end{table*}

As the encoding of quantum information is mode basis-dependent, building on the formalism and measurement possibilities of section \ref{III} we demonstrate the versatility of our platform by presenting two strategies for cluster-state generation: \Final{one where the individual modes are set as the nodes of the cluster, and another based on nonlinear supermodes.} \Da{We exhibit below how optimized configurations can produce different types of clusters shown in Figure \ref{F9} focusing on the individual mode basis, and discuss applications of both encoding schemes in section \ref{VI}.}

\subsection{\Da{Individual modes as nodes of the cluster}} \label{Va}

In the individual mode basis, we set the simplest labelling: \David{the $i$th node of the graph corresponds to the $j$th mode.} The large size of the parameter space corresponding to the ANW enables the optimized generation of different classes of cluster states. We use the sum of the five nullifier variances $F_{C}(\vec{\eta}, \vec{\phi}, \vec{\theta})=\sum_{i=1}^{5} V(\bar{\delta}_{i})$ with 15 free parameters as the fitness function to minimize. For the sake of comparison, we use the same \NewText{parameters} as in the previous section, i.e. a homogeneous coupling profile $f_{j}=1$ with coupling strength $C_{0}=0.24$ mm$^{-1}$ and a fixed length $z=30$ mm. Table \ref{Table1} details the optimized nullifier variances obtained for the $N=5$ nodes cluster states shown in \NewText{Figure \ref{F9}}. Remarkably, we have found realistic \NewText{sets of parameters $\{\vec{\eta}, \vec{\phi}, \vec{\theta}\}$} to generate the clusters in the five analyzed cases. In particular, the linear cluster is optimized with a pump profile very close to the flat one that we introduced in section \ref{III}. \commentsB{We have used here the same set of pump parameters for optimizing star-shaped and GHZ clusters, since both are related by only local oscillator phase shifts \cite{Menicucci2007}.} Full inseparability of the generated states is ensured \NewText{using} the variances of the nullifiers exhibited in Table \ref{Table1} in the expressions shown in appendix \ref{apA}. \comments{We have checked this approach on cluster state generation with up to $N=15$ modes obtaining similar values of nullifier variances for realistic parameters (not shown). Note that the use of a different fitness function or node labelling can lead to different values of nullifier variances.} \David{Therefore, the generation of cluster states in a fixed detection basis is possible in ANWs. Enough degrees of freedom in the pump and the detection are available to compensate for the monolithic coupling structure of the array and yield significant entanglement.}

\subsection{\Da{Cluster states based on nonlinear supermodes}}\label{Vb}

Lower nullifier variances and hence closer-to-ideal cluster states can be obtained using a non-fixed custom detection basis. As introduced in section \ref{III}, the covariance matrix measured at the output of the ANW can indeed be diagonalized by a Bloch-Messiah decomposition through Equation (\ref{BMD}) leading to nonlinear supermodes. Likewise, the covariance matrix associated to a given $N$-mode cluster $V_{C}$ can be constructed from N phase squeezed input states. We write $V_{C}=S_{C} \,\bar{K}^{2}(\vec{r})\, S_{C}^{T}$, where $S_{C}$ is the symplectic and orthogonal transformation --a change of mode basis-- which produces the cluster associated with \MOD{$J$}, and $\bar{K}^{2}(\vec{r})$ is a symplectic diagonal matrix which stands for phase-squeezed states. The elements of $\vec{r}$ are the squeezing parameters corresponding to each mode of encoding. Thus, setting the nonlinear supermodes as the basis the cluster is built from, i.e. $K^{2}(\vec{\eta}, \vec{\phi}, z)=\bar{K}^{2}(\vec{r})$, the cluster and the ANW covariance matrices are related by
\begin{equation} \label{VLO}
V_{C}=  S_{LO}(\vec{\eta},\vec{\phi},\vec{\varphi}, z) V(z) S_{LO}^{T}(\vec{\eta},\vec{\phi},\vec{\varphi}, z), 
\end{equation}
where $S_{LO}(\vec{\eta},\vec{\phi},\vec{\varphi}, z)=S_{C} \,\bar{\mathcal{O}}(\vec{\varphi}) R_{1}^{T} (\vec{\eta},\vec{\phi},z)$ is a symplectic and orthogonal matrix, and $\bar{\mathcal{O}}(\vec{\varphi})$ is an orthogonal matrix related to the freedom of distributing the degree of squeezing among the cluster nodes (see appendix \ref{apB}). Then, the covariance matrix associated to the cluster state can be retrieved from the one related to the ANW via $S_{LO}$.  Below we show two different ways to access cluster states with this encoding.

The first approach is to use a LO with a spatial profile given by the complex representation of $S_{LO}$. \NewText{As discussed \MOD{at section \ref{IIIb}}, this can be carried out by means of a single-mode free space BHD or via a multimode fibered BHD analogously to what was done in the spectral domain \cite{Cai2017}}. \comments{In the present case any cluster is reachable through suitable shaping of a multimode LO with common phase and amplitude references \Final{(see Figure 1)}. Thus, what is to optimize is the distribution of squeezing among the different nonlinear supermodes, i.e. the elements of $K^{2}(\vec{\eta}, \vec{\phi}, z)$ \cite{Arzani2018}. Once the squeezing is optimized, the multimode LO is shaped using the profile of $S_{LO}(\vec{\eta},\vec{\phi},\vec{\varphi}, z)$ \Final{to measure each nullifier.}

The second approach is based on the emulation of the statistics of a given cluster state. This is carried out with a multimode fibered BHD with independent single-mode LOs and postprocessing the photocurrents coming from every detector \cite{Armstrong2012, Ferrini2013}. In this case the transformation $S_{LO}$ can be decomposed as $S_{LO}=\bar{O}_{post}(\vec{\vartheta}) D_{LO}(\vec{\theta})$, where $D_{LO}(\vec{\theta})$ and $\bar{O}_{post}(\vec{\vartheta})$ are transformations associated to the LO phase profile and to the postprocessing gains, respectively (see appendix \ref{apB}).} \David{Note that emulation and LO shaping are not conceptually different, but related to the BHD detection scheme experimentally available.} \Da{We show a detailed example of this approach with the cluster states of Figure \ref{F9} in the appendix \ref{apB}. }

\Final{Thus, the nullifiers associated to any N-dimensional cluster state can be directly measured with a suitable shaped LO via Equation (\ref{VLO}), or approximately emulated by postprocessing}. Moreover, a large class of Gaussian computations can be performed replacing $S_{C}$ by $S_{C}^{'}=U_{comp} S_{C}$, with $U_{comp}$ the orthogonal matrix associated to the required computation \cite{Ferrini2013, Ferrini2016}.

\section{Discussion and perspectives}\label{VI}

\David{We finish first discussing on the encoding and processing of information in different mode bases. We then give perspective on the range of application of ANW-based quantum information processing, and we finally discuss on the feasibility of our approach.}

\David{The individual mode basis is a natural basis to implement cluster states in the spatial domain due to a simple detection scheme \cite{Yukawa2008}. This basis is however harder to implement in other domains, for instance in the spectral domain \cite{Arzani2018}. As we have demonstrated above, the ANW is a versatile source of cluster states with this straightforward encoding. The nonlinear supermode basis offers an even higher flexibility through postprocessing or shaping the basis of detection at the cost of an increased complexity in the detection setup. }

\David{Let us compare both the encoding schemes of quantum information we have introduced above, the individual mode basis and the nonlinear supermode basis, in terms of their respective abilities in quantum information processing and quantum networks --or quantum graphs \cite{Note1}--. In the individual mode basis the measurement basis is fixed. The quantum network is physically yielded by the array with spatially distant nodes, which allows simultaneous access to all the nodes of the network and the detection is addressed by independent BHDs. \Da{See Table \ref{Table2} for a summary of possible applications of the cluster states shown in Figure \ref{F9} or similar}. In contrast, any other basis based on nonlinear supermodes does not have a fixed measurement basis as the supermodes change at every propagation distance. The quantum graph is produced by LO shaping or emulated by postprocessing. Each node is encoded in the whole profile of the output SPDC light, allowing access only to one node of the network at a time.} The resource of both quantum networks and graphs is entanglement. However, multimode entanglement relies on both the quantum state and measurement process \cite{Cai2017}. Thus, even single-mode squeezed states in the individual mode basis encoding can produce quantum graphs in the nonlinear supermode-based encoding with a suitable detection. \Da{This basis has been used to for instance simulate a quantum secret sharing protocol \cite{Cai2017}}.
\begin{table}[]
\centering
\begin{tabular}{l c c}
\hline\hline
Graph & Application & Reference\\[0.5ex]
\hline
Linear & {\centering \hspace{0.5cm}\begin{tabular}{@{}c@{}}Single-mode \\quantum computing\end{tabular}} & \cite{Ukai2010}   \\[1.2ex]
 \begin{tabular}{@{}l@{}}Square\\
(+time multiplexing)\end{tabular} & \hspace{0.5cm} \begin{tabular}{@{}c@{}}Universal \\quantum computing\end{tabular} & \cite{Larsen2019, Asavanant2019} \\[1.2ex]
Star \& GHZ &  \hspace{0.5cm}\begin{tabular}{@{}c@{}}Anonymous \\broadcasting\end{tabular} & \cite{Christandl2005, Menicucci2018}  \\[1.2ex]
Pyramid &  \hspace{0.5cm}\begin{tabular}{@{}c@{}}Quantum \\secret sharing\end{tabular} & \cite{Hillery1999} \\[1.2ex]
Any of the above & \hspace{0.5cm}\begin{tabular}{@{}c@{}}Distributed \\quantum sensing \end{tabular} & \cite{Guo2019} \\[1.2ex]
\hline
\end{tabular}
\caption{\label{Table2}\small{\Da{Some applications of individual mode-based cluster states generated in an ANW.}}}
\end{table}

\David{We assess now the usefulness of both approaches in terms of a specific quantum protocol like MBQC.} The individual mode basis enables sequential or one-shot MBQC. Remarkably, universal single-mode Gaussian operations are possible using linear cluster states, homodyne detection and classical postprocessing \cite{Ukai2010, Su2013}. It is particularly easy in this basis to couple the state to be computed to the cluster by means of a 3 dB fibered beam splitter. Thus the individual mode basis is a convenient basis for \MOD{traditional MBQC. Direct Gaussian MBQC is possible in the nonlinear-supermode postprocessing approach, however it is not universal in general \cite{Ferrini2016}.} Two-dimension cluster states are required for the most general universal MBQC \cite{Menicucci2006}. Recently, the generation of this class of states has been demonstrated by spatial and time-domain multiplexing \cite{Larsen2019, Asavanant2019}. The level of squeezing necessary to implement fault-tolerant MBQC is nevertheless far from technologically available \cite{Menicucci2014}. Fault-tolerant MBQC is potentially realizable with lower squeezing thresholds by using cluster states of higher dimension \cite{Fukui2018}. It is then foreseeable that future MBQC will be based on multiplexing the CV modes in space, time, frequency or angular momentum, and on integration on chip \cite{Pfister2019}. \David{The ANW represents a potential platform to implement that technology as the spatial encoding can be multiplexed in frequency and time in the pulsed regime.}

We would now like to disclose some possible research directions which follow from this work. The first is emulation of quantum complex networks \cite{Nokkala2018}. The dynamics of an ensemble of quantum harmonic oscillators linked according to a specific topology can be mapped to our multimode platform through the symplectic propagator $S(z)$ of section \ref{III}. The temporal evolution of a quantum network $S_{net}(t)$ is directly mapped to our propagator $S(z)$, which can be experimentally realized by adequate pump profile optimization and multimode BHD. The tunability our approach enables the study of different network topologies with a single setup. 

Another interesting feature of integrated ANW is the possibility to include non-Gaussian operations on the quantum state, cornerstone of quantum advantage in CV-MBQC. Single-photon subtraction can be indeed implemented by means of introducing defects in the array, i.e. a weakly coupled single waveguide in between two ANW in such a way that the detection of a single photon de-Gaussifies and entangles the propagating quantum states related to each array \cite{Barral2017b, Walschaers2018}. 

Additionally, besides the practical applications in CV previously discussed such as quantum secret sharing and MBQC, an appealing exploitation of the ANW in DV is Gaussian boson sampling (GBS) \cite{Hamilton2017}. As every $N$-mode Gaussian state generated in the array can be decomposed by Autonne-Takagi (Bloch-Messiah) into $N$ single-mode squeezers in between two linear interferometers, the sampled photon pattern at the output of the ANW enables the computation of the hafnian of the matrix which characterizes the quantum state. \David{Moreover, the ANW operating in the SPDC or optical parametric amplification regimes can be a suitable platform for the simulation of molecular vibronic spectra through GBS \cite{Huh2015, Clements2017, Paesani2018}.} 

Finally, we discuss the feasibility of our approach with respect to the state of the art in integrated CV. \Da{The pump coherence along its profile and with the local oscillator is crucial to the generation and appropriate detection of the generated states. Mechanical and thermal stabilities similar to the ones currently achieved in single PPLN waveguides are expected to be achieved in the experimental implementation of our multiport scheme \cite{Lenzini2018}. Both pump and detection schemes are suitable to be integrated on chip limiting issues associated with pump phase noise \cite{Jin2014, Mondain2019}. Waveguides few microns wide and waveguide spacing on the order of tens of microns are within the scope of standard technology \cite{Alibart2016}. The required homogeneity in nonlinearity and coupling is well within what is routinely achieved by current technology and one of the strengths of our proposal is that many external parameters (pump and LO profile, postprocessing) can compensate for imperfections. Current 127 $\mu$m-spacing commercial V-groove arrays limit the maximum number of waveguides of the ANW. The larger the number of waveguides the longer the chip in order to avoid the losses related to the bends in the input-ouput waveguiding region. With proton-exchange technology we estimate $N \approx$ 8-16 waveguides as reasonable \cite{Alibart2016, Mondain2019}. The limiting factor of integration density can be overcome with nanophotonic nonlinear technologies \cite{Boes2018, Loncar2018}.} The influence of losses on the CV entanglement can be included in our analysis by standard methods \cite{Barral2019c}. Propagation losses have a small impact on squeezing and entanglement assuming typical values in PPLN waveguides ($\approx$ 0.14 dB cm$^{-1}$) \David{and sample lengths (2-3 cm)} \cite{Lenzini2018, Mondain2019}. Nonlinearities as high as $g=24 \times 10^{-4}$ mm$^{-1}$ mW$^{-1/2}$, and coupled cw pump powers ranging from tens to few hundreds milliwatts, have been recently shown in soft proton exchange and ridge PPLN waveguides \cite{Mondain2019, Kashiwazaki2020}. A squeezing level as high as -6.3 dB in cw has been recently demonstrated int a PPLN chip \cite{Kashiwazaki2020}. Furthermore, nanophotonic PPLN waveguides promise to increase the nonlinear efficiency one order of magnitude \cite{Boes2018, Loncar2018}. The pump profile engineering can be realized by means of \Da{telecom} off-the-shelf elements such as fiber attenuators, phase shifters and V-groove arrays, or by means of active elements in electrooptics materials such as LN \cite{Lenzini2018}. \Da{An overall squeezing detection efficiency of 94$\%$ has been recently reported using anti-reflection coating on the chip output facet and balanced-homodyne-detector photodiodes with 99 $\%$ quantum efficiency \cite{Kashiwazaki2020}}. The use of V-groove arrays to fiber the output light can also lead to balanced-homodyne-detector spatial mode-matching visibilities of 99$\%$.

\David{In conclusion, we have demonstrated that the array of nonlinear waveguides is a versatile synthesizer of spatial multimode squeezing and multipartite entanglement. The tuning knobs available in the array of nonlinear waveguides enable a formidable degree of reconfigurability in a compact device. We have in particular introduced a general formalism to analyze the generation of quantum states in nonlinear waveguide arrays. We have shown versatile and significant multipartite entanglement in the experimentally most convenient basis --the individual mode basis--. We have theoretically demonstrated that various cluster states appropriate for quantum networks and measurement-based quantum computing can be generated in arrays of nonlinear waveguides with the same optical setup, using pump and measurement shaping.  Features such as scalability, reconfigurability, subwavelength stability, reproducibility and low cost make this platform an appealing quantum technology in the spatial encoding domain with respect to previous bulk-optics-based multipartite-entanglement approaches. The analysis carried out  here demonstrates that the array of nonlinear waveguides is a competitive contender for quantum communication, quantum computing and quantum simulation.}

\section*{Acknowledgements} The authors thank G. Patera for insightful comments. This work was supported by the Agence Nationale de la Recherche through the INQCA project (Grants No. PN-II-ID-JRP-RO-FR-2014-0013 and No. ANR-14-CE26-0038), the Paris Ile-de-France region in the framework of DIM SIRTEQ through the project ENCORE, and the Investissements d'Avenir program (Labex NanoSaclay, reference ANR-10-LABX-0035).

\appendix

\section{}\label{apA}
Below we define the five normalized nullifiers and four VLF inequalities corresponding to the 5-nodes cluster states exhibited in Figure \ref{F9}. Table \ref{Table1} encodes node $i$ in the individual mode $j$, ($i=j=1, \dots, 5$). \Da{Table \ref{Table3} of appendix \ref{apB}} encodes node $i$ in the nonlinear supermode $m$ ($i=m=1, \dots, 5$). The upper bounds for complete inseparability are slightly different from the usual ones because of the use of normalized nullifiers \cite{Takeda2019, Larsen2019}.

 
a) Linear
\begin{align}\nonumber
\bar{\delta}_{1}&=\frac{y_{1} (\theta_{1}) - x_{2}(\theta_{2})}{\sqrt{2}}, \\  \nonumber
\bar{\delta}_{2}&=\frac{y_{2} (\theta_{2}) - x_{1}(\theta_{1}) - x_{3}(\theta_{3})}{\sqrt{3}}, \\  \nonumber
\bar{\delta}_{3}&=\frac{y_{3} (\theta_{3}) - x_{2}(\theta_{2}) - x_{4}(\theta_{4})}{\sqrt{3}}, \\  \nonumber
\bar{\delta}_{4}&=\frac{y_{4} (\theta_{4}) - x_{3}(\theta_{3}) - x_{5}(\theta_{5})}{\sqrt{3}}, \\  \nonumber
\bar{\delta}_{5}&=\frac{y_{5} (\theta_{5}) - x_{4}(\theta_{4})}{\sqrt{2}}, \\   \nonumber
V(\bar{\delta}_{i})+&V(\bar{\delta}_{i+1})\geq
\begin{cases}   
\sqrt{\frac{8}{3}} \quad &\text{for $i=1, N-1$},\\ 
\quad \frac{4}{3} \quad \quad &\text{for $i=2, \dots, N-2$}.
\end{cases}
\end{align}

b) Pentagon
\begin{align}\nonumber
\bar{\delta}_{i}=&\frac{y_{i} (\theta_{i}) - [x_{i+1}(\theta_{i+1})+x_{i-1}(\theta_{i-1})]}{\sqrt{3}},   \\  \nonumber
&V(\bar{\delta}_{i})+V(\bar{\delta}_{i+1})\geq \frac{4}{3} \quad \text{for}\quad i=1, \dots, 4.
\end{align}
with $x_{0}(\theta_{0})\equiv x_{5}(\theta_{5})$ and $x_{6}(\theta_{6})\equiv x_{1}(\theta_{1})$. 

c) Star
\begin{align}\nonumber
&\bar{\delta}_{i}=\frac{y_{i} (\theta_{i}) - x_{3}(\theta_{3})}{\sqrt{2}} \quad \text{for}\quad i\neq 3, \\  \nonumber
&\bar{\delta}_{3}=\frac{y_{3} (\theta_{3}) - \sum_{i\neq 3}^{5} x_{i}(\theta_{i})}{\sqrt{5}},   \\  \nonumber
&V(\bar{\delta}_{i})+V(\bar{\delta}_{3})\geq \sqrt{\frac{8}{5}}  \quad\quad \text{for}\quad i\neq 3. 
\end{align}

d) Square Pyramid
\begin{align}\nonumber
&\bar{\delta}_{1}=\frac{y_{1} (\theta_{1}) - [x_{2}(\theta_{2})+x_{3}(\theta_{3})+ x_{5}(\theta_{5})]}{2}, \\  \nonumber
&\bar{\delta}_{2}=\frac{y_{2} (\theta_{2}) - [x_{1}(\theta_{1})+x_{3}(\theta_{3})+ x_{4}(\theta_{4})]}{2}, \\  \nonumber
&\bar{\delta}_{3}=\frac{y_{3} (\theta_{3}) - \sum_{i\neq 3}^{5} x_{i}(\theta_{i})}{\sqrt{5}},   \\  \nonumber
&\bar{\delta}_{4}=\frac{y_{4} (\theta_{4}) - y_{1}(\theta_{1})}{\sqrt{2}}, \\  \nonumber
&\bar{\delta}_{5}=\frac{y_{5} (\theta_{5}) - y_{2}(\theta_{2})}{\sqrt{2}}, \\  \nonumber
&V(\bar{\delta}_{i})+V(\bar{\delta}_{3})\geq \sqrt{\frac{8}{5}}  \quad\quad \text{for}\quad i=4, 5.
\end{align}
We have used here the fact that linear combinations of nullifiers are also nullifiers in order to define $\bar{\delta}_{4,5}$. Due to the symmetry of the cluster, the four VLF inequalities are degenerate two by two.

e) GHZ

The GHZ cluster state is equivalent to the star cluster by means of a $\pi/2$ LO rotation for all modes $i\neq3$ as demonstrated in ref. \cite{Menicucci2007}. Applying this labelling we obtain
\begin{align}\nonumber
&\bar{\delta}_{i}=\frac{x_{i} (\theta_{i}) - x_{3}(\theta_{3})}{\sqrt{2}} \quad \text{for}\quad i\neq 3, \\  \nonumber
&\bar{\delta}_{3}=\frac{\sum_{i=1}^{5} y_{i}(\theta_{i})}{\sqrt{5}},   \\  \nonumber
&V(\bar{\delta}_{i})+V(\bar{\delta}_{3})\geq \sqrt{\frac{8}{5}}  \quad\quad \text{for}\quad i\neq 3. 
\end{align}
Thus, a GHZ cluster state is generated using the same set of parameters as that obtained for the star cluster with a $\pi/2$ LO rotation in all the modes except the mode $3$.

\section{}\label{apB}

\begin{table*}[]
\centering
\begin{tabular}{c c c c c c}
\hline\hline
Graph    &   Nullifiers $\{V(\bar{\delta}_{i})\}$  & $\vec{\eta} \times 10^{2}$ mm$^{-1}$ & $\vec{\phi}/ \pi$ & $\vec{\theta}/\pi$   \\[0.5ex]
\hline
Linear & $\{0.29, 0.36, 0.45, 0.13, 0.09\}$ &$\{4.1, 1.8, 3.6, 1.3, 1.0\}$ & $\{0.78, -0.59, -0.26, 0.34, -0.97\}$ & $\{-1.09, -0.09, 0.37, 1.37, 0.24\}$ \\[1ex]
Pentagon & $\{0.28, 0.25, 0.31, 0.17, 0.32\}$ &\{0.7, 2.5, 3.3, 3.4, 1.4\} & \{-0.01, -1.41, -0.40, -0.16, 0.26\} & \{-0.47, -2.57, -0.32, 2.15, 0.72\}\\[1ex]
Star &$\{0.30, 0.28, 0.35, 0.43, 0.20\}$  & \{3.2, 0.5, 0.9, 0.5, 1.7\} & \{-0.02, 0.75, -0.34, -0.54, 0.04\} & \{-0.24, -0.08, 0.15, 0.28, -0.23\}  \\ [1ex] 
Pyramid &$\{0.47, 0.38, 0.19, 0.41, 0.26\}$ &$\{1.3, 0.4, 3.5, 1.7, 3.3\}$ & $\{1.21, 1.01, -0.09, 1.60, -0.08\}$ & $\{0.26, -1.01, -0.09, -0.68, 0.53\}$ \\ [1ex]
GHZ &$\{0.48, 0.31, 0.38, 0.23, 0.24\}$  & \{4.0, 1.5, 2.8, 2.4, 4.3\} & \{0.13, -0.04, -0.08, -0.04, -0.04\} & \{0.56, 0.07, 0.24, 0.08, -0.44\}  \\ [1ex] 
\hline
\end{tabular}
\caption{\label{Table3}\small{\David{Nonlinear supermode basis-based} cluster state generation in a 5-waveguides ANW with homogeneous coupling profile for linear, pentagon, star, square pyramid and GHZ graphs. We show the value of the nullifiers $\{V(\bar{\delta}_{i})\}$, pump power profile $\vec{\eta}$, pump phase profile $\vec{\phi}$ and local oscillator phase profile $\vec{\theta}$. The matrices $\bar{\mathcal{O}}$ and $\bar{\mathcal{O}}_{post}$ obtained for each case are shown in the appendix \ref{apB}. Simultaneous values of $V(\bar{\delta}_{i})$ under the shot noise threshold $V(\bar{\delta}_{i})=1$ are a signature of \NewText{successful} cluster generation. $C_{0}=0.24$ mm$^{-1}$. $z=30$ mm.}}
\end{table*}

\Da{In this appendix we derive Equation (\ref{VLO}), introduce some definitions used in section \ref{V} and show an example of emulation of the statistics associated to the 5-nodes linear cluster state of Figure \ref{F9} using nonlinear supermodes.}

$S_{C}$ is a symplectic and orthogonal transformation which produces the cluster Equation (\ref{Nul}) from N phase ($\hat{y}$)-squeezed input states given by a diagonal matrix with positive entries $\bar{K}^{2}(\vec{r})$. The cluster has thus an associated covariance matrix $V_{C}=S_{C} \,\bar{K}^{2}(\vec{r})\, S_{C}^{T}$. A symmetric cluster transformation $S_{C}$ can be obtained from the adjacency matrix \MOD{$J$} as follows \cite{Ferrini2013}
\begin{equation}\nonumber
S_C=\begin{pmatrix} X_{s}& -Y_{s}\\ Y_{s} & X_{s}\end{pmatrix},
\end{equation}
with \MOD{$X_{s}=(J^{2}+I)^{-1/2}$, $Y_{s}=J X_{s}$} and $I$ the identity matrix. The transformation related to the cluster shape $S_{C}$ is defined up to a transformation $\bar{S}_{C}(\vec{\varphi})=S_{C} \,\bar{\mathcal{O}}(\vec{\varphi})$, with $\bar{\mathcal{O}}(\vec{\varphi})=\diag\{\mathcal{O}(\vec{\varphi}),\mathcal{O}(\vec{\varphi})\}$ and $\mathcal{O}(\vec{\varphi})$ is an N-dimensional orthogonal matrix. This degree of freedom is related to the freedom of distributing the degree of squeezing among the cluster modes $\tilde{K}^{2}(\vec{r},\vec{\varphi})=\bar{\mathcal{O}}(\vec{\varphi}) \bar{K}^{2}(\vec{r})  \bar{\mathcal{O}}(\vec{\varphi})^{T}$. Thus, setting the nonlinear supermodes as the basis the cluster is built from, i.e. taking $K^{2}(\vec{\eta},\vec{\phi}, z)=\tilde{K}^{2}(\vec{r})$, the cluster $V_{C}$ and the ANW $V(z)$ covariance matrices are related through Equation (\ref{BMD}) by
\begin{equation} \nonumber
V_{C}=  S_{LO}(\vec{\eta},\vec{\phi},\vec{\varphi}, z) V(z) S_{LO}^{T}(\vec{\eta},\vec{\phi},\vec{\varphi}, z), 
\end{equation}
where $S_{LO}(\vec{\eta},\vec{\phi},\vec{\varphi}, z)=S_{C} \,\bar{\mathcal{O}}(\vec{\varphi}) R_{1}^{T} (\vec{\eta},\vec{\phi},z)$. Note that we have used here $K^{2}(z) \equiv K^{2}(\vec{\eta},\vec{\phi}, z)$ and  $R_{1}(z) \equiv R_{1}(\vec{\eta},\vec{\phi},z)$ in order to emphasize the dependence of the Bloch-Messiah decomposition with the pump parameters.

The postprocessing matrix $\bar{\mathcal{O}}_{post}(\vartheta)$ is an orthogonal matrix given by $\bar{\mathcal{O}}_{post}(\vartheta)=\diag\{\mathcal{O}_{post}(\vartheta), \mathcal{O}_{post}(\vartheta)\}$. Both matrices ${\mathcal{O}}(\varphi)$ and ${\mathcal{O}}_{post}(\vartheta)$ can be decomposed into $N(N-1)/2$ rotation matrices parametrized for instance by generalized Euler angles \cite{Arfken2005}. In our simulations of 5-node cluster states we have used 10 generalized Euler angles to parametrize each matrix: $\vec{\varphi}=(\varphi_{1}, \dots, \varphi_{10})$, $\vec{\vartheta}=(\vartheta_{1}, \dots, \vartheta_{10})$. These matrices take into account all possible rotations in a 5-dimensional space.

$D_{LO}(\vec{\theta})$ is the matrix related to the LO phase profile $\vec{\theta}$, written as 
\begin{equation}\nonumber
 \quad D_{LO}(\vec{\theta})=\begin{pmatrix} \cos{(\vec{\theta})} & \sin{(\vec{\theta})} \\ -\sin{(\vec{\theta})} & \cos{(\vec{\theta})} \end{pmatrix},
\end{equation}
with $\cos{(\vec{\theta})}=\diag\{\cos{(\theta_{1})}, \dots, \cos{(\theta_{j})}, \dots, \cos{(\theta_{N})}\}$, and $\sin{(\vec{\theta})}$ is defined equally. We have used 5 local oscillator phases $\vec{\theta}=(\theta_{1}, \dots, \theta_{5})$ in our simulations.

\Da{We show an example of the method ii) introduced in section \ref{Vb} with the cluster states of Figure \ref{F9}. The nodes of a given graph are obtained applying the corresponding transformations $\bar{S}_{C}(\vec{\varphi})$ to the nonlinear supermodes. We choose to minimize the following fitness function with 35 free parameters $F_{P}(\vec{\eta},\vec{\phi},\vec{\varphi},\vec{\theta},\vec{\vartheta})=\left\Vert S_{LO}(\vec{\eta},\vec{\phi},\vec{\varphi}, z)-\bar{O}_{post}(\vec{\vartheta}) D_{LO}(\vec{\theta})\right\Vert$, where we use the Frobenius norm ${\small \left\Vert A \right\Vert^{2}=\sum_{i,j} \vert A_{i,j} \vert^{2}}$. The optimization of this function produces a quantum state characterized by $V(z)$, which after BHD in adequate individual quadratures and postprocessing of electrical gains is fully equivalent to a given cluster state characterized by $V_C$. Table \ref{Table3} details the optimized nullifier variances obtained for the $N=5$ nodes cluster states shown in Figure \ref{F9}. We use the same working point as for Table \ref{Table1} above: homogeneous coupling profile $f_{j}=1$ with coupling strength $C_{0}=0.24$ mm$^{-1}$ and a fixed length $z=30$ mm. Remarkably, we find again realistic set of pump and LO parameters $\{\vec{\eta}, \vec{\phi}, \vec{\theta}\}$ where the clusters are generated in the five analyzed cases. The nullifiers thus produced violate also the inseparability conditions shown in the appendix \ref{apA}.}  }

The parameters ($\vec{\eta}, \vec{\phi}, \vec{\theta}$) optimized for each cluster are displayed in Table \ref{Table3}. We show below as an example the matrices $X_{s}$, $Y_{s}$, $\mathcal{O}$ and $\mathcal{O}_{post}$ used to optimize the generation of the linear cluster. Similar values are obtained for the other clusters shown in Table \ref{Table3}.

\begin{equation}\nonumber
X_{s}={\begin{pmatrix}
\frac{(3 \sqrt{2}+5)}{12} & 0 & -\frac{1}{6} & 0 & \frac{(5-3 \sqrt{2})}{12} \\
0 & \frac{(\sqrt{2}+1)}{4} & 0 & \frac{(1-\sqrt{2})}{4} & 0 \\
-\frac{1}{6} & 0 & \frac{2}{3} & 0 & -\frac{1}{6} \\
0 & \frac{(1-\sqrt{2})}{4} & 0 & \frac{ (\sqrt{2}+1)}{4} & 0 \\
\frac{(5-3 \sqrt{2})}{12} & 0 & -\frac{1}{6} & 0 & \frac{(3 \sqrt{2}+5)}{12} \\
\end{pmatrix}},
\end{equation}
\begin{equation}\nonumber
Y_{s}=\\{\begin{pmatrix}
0 & \frac{(\sqrt{2}+1)}{4} & 0 & \frac{(1-\sqrt{2})}{4} & 0 \\
\frac{(\sqrt{2}+1)}{4} & 0 & \frac{1}{2} & 0 & \frac{(1-\sqrt{2})}{4} \\
0 & \frac{1}{2} & 0 & \frac{1}{2} & 0 \\
\frac{(1-\sqrt{2})}{4} & 0 & \frac{1}{2} & 0 & \frac{(\sqrt{2}+1)}{4} \\
0 & \frac{(1-\sqrt{2})}{4} & 0 & \frac{(\sqrt{2}+1)}{4} & 0 \\
\end{pmatrix}},
\end{equation}
\begin{equation}\nonumber
{\mathcal{O}}={\begin{pmatrix}
-0.03 & -0.61& -0.62 & 0.40 & 0.29 \\
0.02 & 0.28 & -0.38 & -0.63 & 0.62 \\
0.01 & 0.21 & 0.45 & 0.51 & 0.70 \\
0.32 & 0.67 & -0.49 & 0.41 & -0.19 \\
-0.95 & 0.25 & -0.14 & 0.12 & -0.06
\end{pmatrix}},
\end{equation}
\begin{equation}\nonumber
{\mathcal{O}}_{post}={\begin{pmatrix} 
0.11 & 0.62 & -0.62 & -0.29 & 0.36 \\
0.07 & -0.14 & -0.62 & 0.71 & -0.28 \\
0.73 & -0.18 & -0.14 & -0.42 & -0.49 \\
-0.60 & -0.42 & -0.43 & -0.48 & -0.20 \\
0.30 & -0.62 & -0.15 & 0.00(4) & 0.71
\end{pmatrix}}.
\end{equation}

\end{document}